\documentclass[12pt,preprint]{aastex}

\newcommand{\VEV}[1]{\langle#1\rangle}

\shorttitle{WMAP power asymmetries}
\shortauthors{Hansen,Banday,Gorski,Eriksen,Lilje}

\usepackage{epsfig}

\begin{document}

\title{Power Asymmetry in Cosmic Microwave Background Fluctuations from Full Sky
to Sub-degree Scales: Is the Universe Isotropic?}

\author{F. K. Hansen}
\affil{Institute of Theoretical Astrophysics, University of Oslo, P.O. Box 1029 Blindern, N-0315 Oslo, Norway; \\Centre of Mathematics for Applications, University of Oslo, P.O. Box 1053 Blindern, N-0316 Oslo }
\email{frodekh@astro.uio.no}

\author{A. J. Banday}
\affil{Max Planck Institut fur Astrophysik, Karl Schwarzschild-Str. 1, Postfach 1317, D-85741 Garching bei Munchen, Germany; \\Centre d'Etude Spatiale des Rayonnements, 9 av du Colonel Roche, BP 44346, 31028 Toulouse Cedex 4, France}
\email{banday@MPA-Garching.MPG.DE}

\author{K. M. G\'orski}
\affil{Jet Propulsion Laboratory, M/S 169/327, 4800 Oak Grove Drive, Pasadena CA 91109; \\Warsaw University Observatory, Aleje Ujazdowskie 4, 00-478 Warszawa, Poland; \\California Institute of Technology, Pasadena CA 91125}
\email{Krzysztof.M.Gorski@jpl.nasa.gov}

\author{H. K. Eriksen}
\affil{Institute of Theoretical Astrophysics, University of Oslo, P. O. Box 1029 Blindern, N-0315 Oslo, Norway; \\Centre of Mathematics for Applications, University of Oslo, P.O. Box 1053 Blindern, N-0316 Oslo} 
\email{h.k.k.eriksen@astro.uio.no}

\author{P. B. Lilje}
\affil{Institute of Theoretical Astrophysics, University of Oslo, P. O. Box 1029 Blindern, N-0315 Oslo, Norway; \\Centre of Mathematics for Applications, University of Oslo, P.O. Box 1053 Blindern, N-0316 Oslo}
\email{lilje@astro.uio.no}

\newpage

\begin{abstract}
We repeat and extend the analysis of Eriksen et al 2004 and Hansen et al 2004 testing the isotropy of the Cosmic Microwave Background (CMB) fluctuations. We find that the hemispherical power asymmetry previously reported for the largest scales $\ell=2-40$ extend to much smaller scales. In fact, for the full multipole range $\ell=2-600$, significantly more power is found in the hemisphere centered at $(\theta=107^\circ\pm10^\circ,\phi=226^\circ\pm10^\circ)$ in galactic co-latitude and longitude than in the opposite hemisphere consistent with the previously detected direction of asymmetry for $\ell=2-40$. We adopt a model selection test where the direction and amplitude of asymmetry as well as the multipole range are free parameters. A model with an asymmetric distribution of power for $\ell=2-600$ is found to be preferred over the isotropic model at the $0.4\%$ significance level taking into account the additional parameters required to describe it. A similar direction of asymmetry is found independently in all six subranges of 100 multipoles between $\ell=2-600$ and none of our 9800 isotropic simulated maps show a similarly consistent direction of asymmetry over such a large multipole range. No known systematic effects or foregrounds are found to be able to explain the asymmetry.
\end{abstract}


\section{Introduction}
\label{sect:intro}

In the first public release of the WMAP data, we reported a significant asymmetry of the distribution of large scale power on the sky\citep{asymm1,asymm2}. This finding as well as other statistical anomalies were confirmed by several other authors using other methods \citep{alignment,ilc2,park,vielva,minkowski,axisofevil,bispectrum,spots1,directionalw,curvature,parameters}. Some of these have been confirmed again in the 3 and 5 year data \citep{wmap32,wmap34,cruz,directionalw2,eriksenmod,pietrobon}. In \citep{asymm2} we found that the independent multipole ranges $\ell=2-19$ and $\ell=20-40$ were both particularly asymmetric but with two different axes of asymmetry, the former being a galactic north-south asymmetry and the latter being an east-west asymmetry. Using the full range $\ell=2-40$ we found the highest  significance of the asymmetry with the axis pointing in the direction of $(100^\circ,237^\circ)$ in galactic co-latitude and longitude (this is the convention we will use for all sky positions in this paper). 

The position of the non-Gaussian cold spot found in \citep{vielva} is found to be positioned close to the center of the hemisphere with large fluctuation power. Furthermore, the CMB signal also demonstrated non-Gaussian statistical properties in the direction where the power spectrum amplitude was low \citep{park,minkowski,curvature}. Our own attempts to provide an explanation to the asymmetry invoked a class of homogeneous models that include anisotropic expansion (shear) and global rotation (vorticity) - the Bianchi type VIIh models\citep{tess1,tess2}. Unfortunately, this fails as a physical explanation\citep{tess3,bridges} since the best-fit parameter space is then inconsistent with a wide range of other evidence.

Recent theoretical developments\citep{ackerman,erickcek,gordon} have proposed new mechanisms for generating the imprint of a preferred direction in the CMB. Some of these models would lead to hemispherical asymmetry also for multipoles much larger than $\ell=40$ (angular scales smaller than 5 degrees). A recent analysis\citep{nicolaas} of the Ackerman et al. model indicates a good fit to the WMAP data up to a multipole moment $\ell=400$.  However, this model is not able to address the issue of the hemispherical power asymmetry.

In this paper we reinvestigate the nature of the power asymmetry using the recent 5 year data release from WMAP \citep{wmap5}, and specifically assess whether the signal manifests itself on smaller angular scales $\ell>>40$. Instead of randomly searching for an asymmetry in arbitrary multipole ranges, we adopt a model selection procedure that searches for the best fit asymmetry direction and multipole interval. An asymmetric model of the CMB needs more free parameters (i.e. direction of the axis of asymmetry) than a simpler isotropic model. The model selection procedure tests whether an asymmetric model of the CMB is actually preferred by the data taking into account the introduction of additional free parameters. We introduce a set of general asymmetric models with three to five additional parameters. These models are not based on any physical theory but are rather general parametrization of hemispherical asymmetry. We use these models in order to investigate how the asymmetry is distributed in harmonic space as well as on the sphere.

In section \ref{sect:data} we describe the data and masks used in the analysis. The methods used to assess the asymmetry are outlined in section \ref{sect:method}. In section \ref{sect:res} we show the results of the isotropy tests applied to the 5 year WMAP data and in section \ref{sect:conclusions} we conclude.

\section{Data}
\label{sect:data}

 The analysis in this paper was performed using the 5 year release of the WMAP data (publicly available at the Lambda web site\footnote{http://lambda.gsfc.nasa.gov/}) as well as a large ensemble of simulated maps of each channel Q (41GHz), V (61GHz) and W (94GHz) (the result for each channel is obtained by taking the mean of all DAs for each channel). From all maps, we have subtracted the best fit mono- and dipole. Some tests are also performed on maps from single DAs and single years of observation.

A series of galactic masks are used in the analysis:
\begin{itemize}
\item  {\bf KQ85}: The WMAP KQ85 cut with point source mask. Sky fraction used: 82\%
\item  {\bf KQ75}: The WMAP KQ75 cut with point source mask. Sky fraction used: 72\%
\item  {\bf KQ75 ext.}: The KQ75 cut extended with 5 degrees along the rim of the galaxy. Point source mask unchanged. Sky fraction used: 63\%
\item {\bf $|b|>30$ (sometimes referred to as the 60 degree cut)}: Same as the extended KQ75 mask, but with an additional 60 degree band cut on the galactic equator. Sky fraction used: 47\%.
\item {\bf KQ85N} equals the $|b|>30$ cut in the northern galactic hemisphere and the KQ85 cut in the southern galactic hemisphere. Sky fraction used: 65\%
\item {\bf KQ85S} equals the $|b|>30$ cut in the southern galactic hemisphere and the KQ85 cut in the northern galactic hemisphere. Sky fraction used: 64\%
\end{itemize}

\section{Methodology: Power spectrum estimation on hemispheres}
\label{sect:method}

\subsection{Hemisphere spectra}
\label{sect:met1}

In \citep{asymm1,asymm2}, we estimated the power spectrum on hemispheres centered on 164 different positions on the sphere. In that analysis, we applied the Gabor transform approach \citep{gabor1,gabor2}. In order to speed up the analysis without significant loss of precision, we will here apply the much faster MASTER algorithm \citep{master}. This allows the analysis of many more positions on the sphere as well as a large number of simulations for each WMAP year of observation and each channel.

The power spectrum is estimated from the pseudo power spectrum by \citep{master}
\[
C_\ell=\sum_{\ell'}K^{-1}_{\ell\ell'}(\tilde C_{\ell'}-N_{\ell'})\ \ \ \ \tilde C_\ell=\sum_{m=-\ell}^{\ell}\frac{\tilde a_{\ell m}^X\tilde a_{\ell m}^Y}{2\ell+1}
\]
where $\tilde a_{\ell m}^X$ is the spherical harmonic transform of channel $X$ with a given mask, $\tilde C_\ell$ is the corresponding pseudo power spectrum and $N_\ell$ is the noise power spectrum. The coupling kernel $K_{\ell\ell'}$ depends on the mask applied to the data as detailed in \citep{master}. In previous works we only used the auto-spectra, i.e. $X=Y$. Here we will also use the cross power spectra $X\ne Y$ as an additional check. In the previous papers, we estimated the power spectrum on hemispheres centered on 164 different positions on the sphere. Here we use the positions of the 3072 pixel centers in a HEALPix\footnote{http://healpix.jpl.nasa.gov/} $N_\mathrm{side}=16$ map. 

Using the above approach, we obtain for each multipole bin $b$ a $N_\mathrm{side}=16$ map $M_i(b)$ where the value of each pixel $i$ corresponds to the $\ell(\ell+1)C_\ell$ power on a hemisphere centered on that pixel. In harmonic space, we bin the spectrum in 2 multipoles per bin such that $\ell_f=2b+2$ and $\ell_l=2b+3$ are the first and last multipoles of bin $b$. We have also performed tests on more localized spectra, i.e. spectra estimated on disks of various sizes:
\begin{itemize}
\item Hemispheres (diameter $180^\circ$), 2 multipoles per bin, $\ell_f=2b+2$ and $\ell_l=2b+3$
\item $90^\circ$ diameter disks, 4 multipoles per bin, $\ell_f=4b+2$ and $\ell_l=4b+5$
\item $45^\circ$ diameter disks, 16 multipoles per bin, $\ell_f=16b+2$ and $\ell_l=16b+17$
\item $22.5^\circ$ diameter disks, 16 multipoles per bin, $\ell_f=16b+2$ and $\ell_l=16b+17$
\end{itemize}

In analogy to the tests considered in the previous papers \citep{asymm1,asymm2}, we tested the asymmetry for different multipole ranges constructing the following map:
\[
M_i^\mathrm{lmin,lmax}=\sum_bM_i^b,
\]
The max power spectrum ratio for a given multipole range was then defined as
\[
r^\mathrm{lmin,lmax}=\mathrm{max}(\frac{M_i^\mathrm{lmin,lmax}}{M_j^\mathrm{lmin,lmax}}),
\]
where $j$ is the pixel opposite to $i$.

Using the multipole ranges $\ell=2-19$ and $\ell=2-41$ for which a significant asymmetry was found in previous papers, we still find a significant asymmetry with a similar direction and significance. 

The problem with this approach however, is that the max asymmetry axis as well as the significance is different for different multipole ranges.  We made some attempts to determine whether the asymmetry continues to higher multipoles and if so, what is the highest multipole where the asymmetry is present.  This turned out to be difficult because of the instability of significances and directions as a function of the multipole range. Whether we have a significant asymmetry or not depends on which multipole range we chose to look at. Also the axis of asymmetry is slightly different for different multipole ranges. 

We will now present an approach which, first of all will solve the problem of choosing which multipole ranges to look at, and secondly will tell us to which degree a complicated asymmetric model is preferred by the data rather than the isotropic model. The approach is inspired by a similar idea in \cite{landmod}. Then in order to understand the results from the model selection method,  we will present a second much simpler asymmetry test which is based on testing the alignment of the axes of asymmetry between independent multipole ranges.

\subsection{Model selection method}
\label{sect:met2}

We can look at the asymmetry in the following way: The asymmetry is the result of a dipole component in the maps $M_i^b$ which is common for several multipole bins $b$. The dipole for a given multipole bin $b$ is given by
\[
a_{1m}^b=\sum_iM_i^bY_{1m}^i.
\]
We thus propose the following asymmetric model:
\[
a_{1m}^b=a_{1m}^b(0)+A(b)a_{1m}(\theta,\phi)
\]
where $a_{1m}^b(0)$ is the random dipole expected in an isotropic model and $a_{1m}(\theta,\phi)$ is the common dipole component. The parameters $(\phi,\theta)$ are assumed to be independent of the bin. We will allow variations of the asymmetry amplitude $A$ with bin. To test this hypothesis, we will apply a simple $\chi^2$ fit,
\begin{equation}
\label{eq:chi2}
\chi^2=\mathbf{d}^\dagger\mathbf{C}^{-1}\mathbf{d},
\end{equation}
where the elements of the data vector $\mathbf{d}$ are given by $d_{mb}=a_{1m}^b(obs)-A(b)a_{1m}(\theta,\phi)$ for all $m=[-1,1]$ and all bins $b$. Here $a_{1m}^b(obs)$ is the dipole of the map $M_i^b$ for the data to be tested. The elements of the correlation matrix are simply $C_{mb,m'b'}=\VEV{a_{1m}^b(a_{1m'}^{b'})^*}$. 
We will minimize this $\chi^2$ with respect to the parameters $(A(b),\theta,\phi)$. The resulting asymmetry direction is the best fit common dipole component for all bins included in the analysis. The value of the $\chi^2$ at the minimum will be compared to the $\chi^2$ for the isotropic hypothesis,
\begin{equation}
\label{eq:chi20}
\chi^2_0=\mathbf{d}^\dagger_0\mathbf{C}^{-1}\mathbf{d}_0,
\end{equation}
where the elements of $\mathbf{d}_0$ are given only by the observed dipole $d_{0,mb}=a_{1m}^b(obs)$.
The $\chi^2$ improvement $\Delta\chi^2=\chi^2_0-\chi^2$ due to the additional parameters will be calibrated with a set of isotropic gaussian simulations. These simulations will give us the expected improvement $\Delta\chi^2$ from the additional parameters to which the $\chi^2$ improvement in the data can be compared. If the improvement $\Delta\chi^2$ in the data turns out to be significantly better than in simulations, this would indicate that the asymmetric model is preferred by the data.

We will test three different models for the asymmetry amplitude $A(b)$,
\begin{enumerate}
\item {Constant amplitude:} We will test (a) a three-parameter model $(A_0,\theta,\phi)$ with constant amplitude $A_0$ for a given set of  multipole ranges, (b) a four-parameter model where the asymmetry is assumed to have the constant amplitude $A_0$ starting at $\ell=2$ until $\ell_{max}$ where the latter is the fourth free parameter, (c) a five-parameter model where both the lowest and highest multipoles in the asymmetric multipole range are free parameters.
\item {Linearly decreasing amplitude:} In this four-parameter model we will assume the asymmetry to be maximal at $\ell=2$ and decrease linearly according to $A=A_0(1-\alpha\ell)$ where $\alpha$ and $A_0$ are free parameters.
\item {Gaussian multipole dependence:} We will test a model where the asymmetry peaks at a certain multipole $\ell_0$ and falls off to both sides following a Gaussian, $A=A_0e^{-(\ell-\ell_0)^2/(2\sigma^2)}$. We will test (a) a four-parameter model which is assumed to peak at $\ell_0=2$ with $\sigma$ as free parameter and a five-parameter model where both $\ell_0$ and $\sigma$ are free parameters.
\end{enumerate}

Before presenting the results, we will show the procedure that we use to obtain these results. The following is the procedure that we use in the case of a 5 parameter model with a flat amplitude $A_0$ (i.e. the amplitude is constant over a multipole range between $\ell_\mathrm{min}$ and $\ell_\mathrm{max}$ and zero for all other multipoles). The free parameters in this model are thus: the two direction angles, the amplitude of the common dipole and the minimum and maximum multipoles of the multipole range where a common dipole component is found (i.e. where the amplitude is nonzero). The simulations are made with a maximum multipole of $L_\mathrm{max}$, so the maximum and minimum multipoles will be sought within the multipoles available from the simulation in the range from $L_\mathrm{min}=2$ to $L_\mathrm{max}$. Since we will be looking for asymmetry extending over large multipole ranges, we will only look for multipole ranges with a minimum number of $\Delta\ell=\ell_\mathrm{max}-\ell_\mathrm{min}$ multipoles with a common dipole.
\begin{enumerate}
\item We make two sets of 1400 WMAP simulations of a given band with a given mask.
\item In each simulation, we estimate the power spectrum $\hat C_b(i)$ for a hemisphere centered on the center of each pixel $i$ in the $N_\mathrm{side}=16$ HEALPix grid. We thus obtain 3072 power spectra, one for each direction on the sky up to a multipole $\ell=L_\mathrm{max}$. For each simulations we thus have one map $M_i^b$ for each bin $b$.
\item We extract the dipole $a_{1m}^b$ from the map $M_i^b$ for each bin $b$ and each simulation.
\item We use the first set of simulations to construct the correlation matrix $\mathbf{C}$.
\item We will now use the second set of simulations: For each simulation, we test all different anisotropic models with different values for the five free parameters, $\ell_\mathrm{min}$, $\ell_\mathrm{max}$, dipole amplitude $A_0$ and direction $(\theta,\phi)$. We will calculate the $\chi^2$ (equation \ref{eq:chi2}) of the simulated map for all these models and record the parameters of the model with the lowest $\chi^2$. We use the following procedure to minimize the $\chi^2$ in each simulation:
\begin{enumerate}
\item We start by making a loop over all possible multipole ranges between $L_\mathrm{min}=2$ and $L_\mathrm{max}$ which are the multipoles which we have available. If for instance $L_\mathrm{max}=300$ and $\Delta\ell=200$, we will go through all models with $(\ell_\mathrm{min}=2,\ell_\mathrm{max}=201)$, $(\ell_\mathrm{min}=2,\ell_\mathrm{max}=221)$, $(\ell_\mathrm{min}=2,\ell_\mathrm{max}=241)$, $...$, $(\ell_\mathrm{min}=21,\ell_\mathrm{max}=221)$, $(\ell_\mathrm{min}=21,\ell_\mathrm{max}=241)$, $...$,  $(\ell_\mathrm{min}=101,\ell_\mathrm{max}=300)$.
\item For each multipole range $(\ell_\mathrm{min},\ell_\mathrm{max})$, we make a loop over a large set of possible values for the model angle $\theta$.
\item For each combination of $\ell_\mathrm{min}$, $\ell_\mathrm{max}$ and $\theta$, we minimize the $\chi^2$ analytically with respect to the model parameters $\phi$ and $A_0$. Having found the values $\phi$ and $A_0$ that minimize the $\chi^2$ we now record the value of the minimum $\chi^2$ given these three values of the model parameters $(\ell_\mathrm{min},\ell_\mathrm{max},\theta)$ as well as the best fit values $A_0$ and $\phi$ for this combination of $(\ell_\mathrm{min},\ell_\mathrm{max},\theta$).
\item After terminating the loop over multipole ranges and directions $\theta$, we now have a three dimensional array of minimum $\chi^2$ values. We search for the minimum value of $\chi^2$ in this array and thereby obtain the set of model parameters $(\ell_\mathrm{min}, \ell_\mathrm{max}, A_0, \theta,\phi)$ which minimizes the $\chi^2$. We now compare this minimum $\chi^2$ with the $\chi_0^2$ (equation \ref{eq:chi2}) obtained assuming the isotropic model. This difference $\Delta\chi^2=\chi_0^2-\chi^2$ is recorded for the given simulation. This value shows the improvement in $\chi^2$ for this given simulation when using a five-parameter anisotropic model.
\end{enumerate}
\item After repeating the above procedure for all 1400 simulations, we repeat the same procedure on the WMAP data. We now have an array of $\chi^2$ improvements from a five-parameter model from all of the simulations as well as for the data. We now check the $\chi^2$ improvement of the data with respect to the simulations. If we quote a significance of $1\%$ it means that $1\%$ of the simulations had a similar or larger improvement of the $\chi^2$ using the five-parameter model. In the results we will also list the best fit multipole range $(\ell_\mathrm{min}, \ell_\mathrm{max})$ as well as the amplitude $A$ and direction $(\theta,\phi)$ of the common dipole in the data.
\end{enumerate}

The exact procedure described above turned out to have convergence problems. The number of simulations used to obtain the correlation matrix was too small and caused small instabilities in the best fit parameters.

The number of simulations necessary to get a converged covariance matrix would be too CPU demanding and we therefore chose the following solutions: By increasing the size of the power spectrum bins, the correlations between bins get smaller. Using power spectrum bins of $\Delta\ell=20$, we found that the correlations can be ignored and only the diagonal part of the correlation matrix is included. With this bin size, identical results are obtained no matter whether the full correlation matrix or only the diagonal is used. With smaller bin sizes there was still some dependence on whether the full matrix was used or not. In the rest of this paper, we will therefore only use power spectra averaged in bins of 20 multipoles and a diagonal correlation matrix. Only in exceptional cases where we look at small multipole ranges will the bins of two multipoles be used. In this case, this will be clearly stated.

\begin{figure}
\begin{center}
\includegraphics[width=0.45\linewidth]{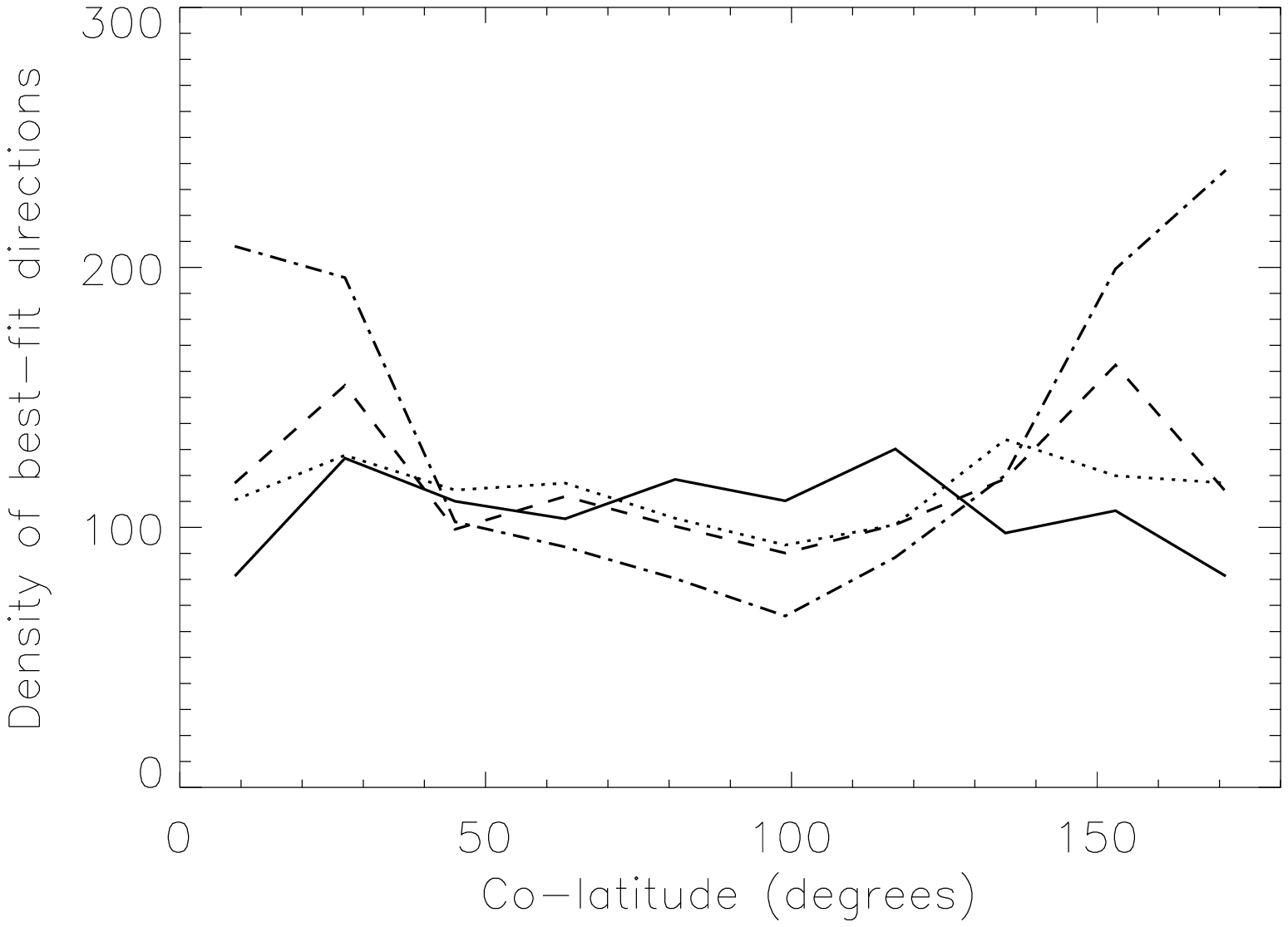}
\caption{The preferred direction for 1400 isotropic simulations using the KQ85 cut (solid line), KQ75 cut (dotted line), extended KQ75 cut (dashed line) and the 60 degree cut (dot-dashed line). Here, the amplitude, direction and $\ell_\mathrm{max}$ were free parameters and may thus be different for each simulation. \label{fig:prefdir}}
\end{center}
\end{figure}

 In figure \ref{fig:prefdir} we show the preferred direction for 1400 simulated maps using various galactic cuts. The plot shows the density of best-fit directions as a function of galactic co-latitude. The KQ85 and KQ75 cut show a uniform distribution of best-fit directions whereas for the larger cuts there is a preference for the poles. Thus we would expect that the preferred direction of asymmetry will be shifted away from the galactic plane for large sky cuts. Because of the large cut, power spectra estimated on hemispheres centered close to the poles will be similar in the polar area. This produces a dipolar structure with an axis pointing towards the poles.

\subsection{Test of alignment of multipole ranges}
\label{sect:met3}

In the results section we will see, using the model selection method described above, that a model with a common dipole component in the range from $\ell=2-600$ is preferred by the data with high significance. There should thus be a strong correlation between the distribution of power in different independent multipole ranges between $\ell=2-600$. The purpose of the alignment test is to check if the asymmetry is distributed over the full multipole range and thus showing up as an alignment of the dipole of the power distribution of different independent multipole ranges.

The idea for the alignment test is simple: we construct the maps
\[
M_i(b1,b2)=\sum_{b=b1}^{b2}M_i(b),
\]
where we sum over (A) blocks of 20 multipoles and (B) blocks of 100 multipoles. Thus for test A, we obtain a set of maps $M_i(2,21)$, $M_i(22,41)$, $M_i(42,61)$ etc. and for test B we obtain $M_i(2,101)$, $M_i(102,201)$, $M_i(202,301)$ etc. (due to different binning for 45 and 22.5 degree tests, the ranges in test B will be as given in table \ref{tab:chires4}).

For each map $M_i(b1,b2)$ the dipole is extracted (by a simple spherical harmonic transform on the map $M_i(b1,b2)$) and the direction of the dipole is stored in a vector $\vec{v}_i$ where $i$ is a multipole range $(b1,b2)$. In order to assess whether these directions for different multipole ranges are significantly more aligned in the WMAP data than in isotropic simulations, we define the mean angular distance $\bar{\theta}$ as
\[
\bar{\theta}=\sum_{ij}\arccos\vec{v}_i\cdot\vec{v}_j,
\]
where the sum over $i$ and $j$ is over subranges$(b1,b2)$ up to the maximum multipole for the given case. We will in the following quantify the alignment of the power distribution in the WMAP data by specifying the number of simulations with a lower mean angular distance $\bar{\theta}$ between the dipoles of the power distribution.

\section{Results}
\label{sect:res}

\subsection{Results with model selection}

\begin{figure}
\begin{center}
\vspace*{-14cm}\hspace*{-5cm}\includegraphics[width=1.6\linewidth]{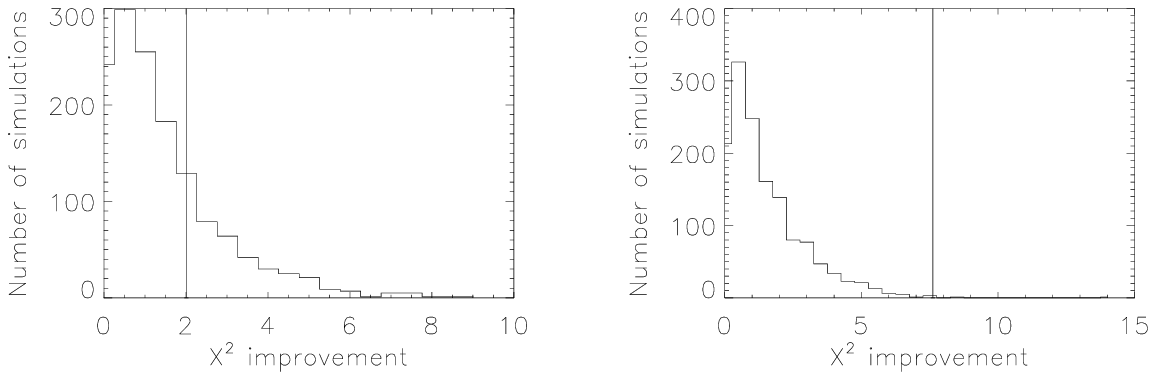}
\vspace*{-17cm}\caption{Histogram of the improvements in $\chi^2$ for a three-parameter model with $(\theta, \phi, A_0)$ as free parameters for 1400 simulated maps using the WMAP V-band parameters with the KQ85 galactic cut. The left plot is for a model with asymmetry in the range $\ell=2-19$, the right plot is for a model with asymmetry in the range $\ell=2-41$. The horizontal line is the improvement in the $\chi^2$ for the 5 year WMAP data. \label{fig:chi2}}
\end{center}
\end{figure}

As a first test of the model selection approach, we used the constant amplitude model with $\ell_\mathrm{min}$ and $\ell_\mathrm{max}$ fixed at the ranges $\ell=2-19$ and $\ell=2-41$, allowing only the amplitude and direction of asymmetry to vary. In this case, we found the best fit asymmetry axis to be $(138^\circ,220^\circ)$ and $(108^\circ,227^\circ)$ being within $10^\circ-15^\circ$ of the asymmetry axis found in \citep{asymm2} obtained with the method described in section \ref{sect:met1}. Note that due to the limited multipole range, we use power spectrum bins of two multipoles in this case, ignoring correlations between multipoles. In figure \ref{fig:chi2} we show the distribution of $\chi^2$ improvements $\chi^2(isotropic)-\chi^2(A_0,\phi,\theta)$ with the three parameters $(A_0,\phi,\theta)$ obtained with 1400 isotropic simulated maps. The vertical line in the plots show $\Delta\chi^2$ for the data. For $\ell=2-41$ we see that only $0.4\%$ of the simulations have a drop in $\chi^2$ similar to the drop seen in the data, showing that the anisotropic model is actually preferred by the data. For the multipole range $\ell=2-19$ however, $30\%$ of the simulations show a similar drop in $\chi^2$ and this asymmetry is therefore not significant alone taking into account the additional number of parameters required to describe it.

We now allow first $\ell_\mathrm{max}$ (fixing $\ell_\mathrm{min}=2$) and then later also $\ell_\mathrm{min}$ to be free parameters, i.e. we set the amplitude $A$ to a constant value $A_0$ in a multipole range ($\ell_\mathrm{min},\ell_\mathrm{max})$ and to zero for all other multipoles. Again we measured the drop in $\chi^2$ by the addition of 4 (fixing $\ell_\mathrm{min}=2$ and varying $\ell_\mathrm{max}$) and 5 parameters (varying both $\ell_\mathrm{min}$ and $\ell_\mathrm{max}$) for the simulations and compared to the data. 

In table \ref{tab:chires2} we show the results for larger scales. We extracted power spectra up to $L_{max}=300$ from WMAP simulation of different channels with different galactic cuts.  As we are looking for asymmetric models extending over a large range in multipole space, we restrict our model search to models with at least 10 consecutive multipole bins (200 multipoles) with a common dipole component. The results for the 4-parameter fits (The four free parameters are the constant amplitude $A_0$, the direction ($\theta,\phi$) and the maximum multipole of asymmetry $\ell_\mathrm{max}$) are presented in the table. 

The table shows that within the multipole range $\ell=2-300$, there is a dipole component with common amplitude and direction for the multipole range $\ell=2-221$. For the KQ85 cut, only $0.1-0.3\%$ of the simulations show a similarly strong fit (similarly large $\chi^2$ improvement) for an asymmetric model. We see that the result is stable with frequency channel. The significance is dropping with larger galactic cuts, but even with the extended KQ75 cut, only $1.5\%$ of the simulations give a similarly strong fit to an asymmetric model. Even with the largest cut, the direction is remarkably consistent, but note the expected shift away from the galactic plane as discussed above. The dipole fitting procedure has thus revealed that the hemispherical power asymmetry extends to at least $\ell=221$.

The fact that the range $\ell=2-221$ is found to be the best fit asymmetric range, does not mean that the asymmetry cannot extend beyond $\ell=221$.  Our model consists of an isotropic field which in our analysis is considered noise and an anisotropic dipole component which is considered the signal. The isotropic 'noise' component to the dipole will randomly change the direction and amplitude of each single power spectrum bin away from the asymmetric direction. By including larger multipole ranges, this noise component is reduced and asymmetries extending over larger multipole ranges even beyond $\ell=300$ can give a good fit. 

From table \ref{tab:chires2} we also see that the best fit asymmetric multipole range is slightly larger for the larger galactic cuts. For KQ85 $\ell=2-221$ is the best fit asymmetric range whereas for KQ75 $\ell=2-281$ gives a better fit and for larger cuts even $\ell=2-300$ is preferred. As discussed above, the fact that a smaller range $\ell=2-221$ gives the best fit for KQ85 does not mean that the larger range $\ell=2-281$ is a bad fit for this mask. In the table we have included in parenthesis the significance for $\ell=2-281$ for KQ85. Clearly $\ell=2-281$ is also significantly asymmetric for KQ85, but because of the random noise component, $\ell=2-221$ gives a slightly better fit.

Note that we have not included the results of the 5-parameter fits. The reason for this is that the 5-parameter fit in all these cases show identical best-fit model parameters (that is, they show that the best fit value for the fifth parameter $\ell_\mathrm{min}$ equals $\ell_\mathrm{min}=2$ for all cases) to the 4-parameter fits. The significances for the 5-parameter model are generally lower than the 4-parameter model even if the best fit parameters are the same. The reason for this is that significances for the 4-parameter model is calibrated with $\chi^2$ improvements in 4-parameter fits to isotropic simulations whereas the 5-parameter model is calibrated with $\chi^2$ improvements in 5-parameter fits to simulations. The simulations will in general have a larger $\chi^2$ improvement for a 5-parameter model than a 4-parameter model. The data however will have the same $\chi^2$ for a model with exactly the same parameters. There will therefore be more simulations with a larger $\chi^2$ improvement for the 5-parameter fit than for the 4-parameter fit even if the $\chi^2$ of the data is the same.

To investigate properly the maximum multipole for asymmetry, we run a set of simulations with $L_\mathrm{max}=500$ and $L_\mathrm{max}=800$. In order to reduce the CPU time for the $L_\mathrm{max}=800$ case, we now use disks of 90, 45 and 22.5 degree diameter. In this way we also obtain more localized spectra. For the 45 and 22.5 degree disks, we needed to use multipole bins of 16 multipoles in the power spectrum estimation, instead of 20 used in the previous analysis. For the 22.5 degree disks, the variance of the power spectrum estimate close to the galactic plane was so large that the map $M_i^b$ needed to be normalized by its standard deviation (obtained from simulations) before a dipole fit could be performed. As a result, the values for the amplitude obtained in this case is different from the amplitudes obtained for other disk sizes.

The results are shown in table \ref{tab:chires3}. The hemisphere results with $L_\mathrm{max}=500$ show that the asymmetry extends to $\ell=481$ with a similar direction of asymmetry as for lower multipoles but now with a lower significance ($p=2.3\%$). For the more localized spectra however, the asymmetry is found highly significant ($p=0.4\%$ for the most localized spectra) for the range $\ell=2-600$ for all disk sizes, but no evidence is found for an asymmetry extending beyond $\ell=600$. In the same table we also show results for the dipole fit on smaller disks including only multipoles up to $L_\mathrm{max}=300$ in order to check for consistency with the hemisphere results. Clearly a consistent dipole fit is found also in this case with more localized power spectra. 

Finally we will make a consistency check by performing the 5-parameter dipole fit in individual subranges of 100 multipoles from $\ell=2$ to $\ell=800$ using 22.5 degree disks. Note that the size of these subranges is not exactly 100 because each power spectrum bin has 16 multipoles. In table \ref{tab:chires4} we show the results. We see clearly that the best fit direction in each subrange up to $\ell=600$ is consistent with the best fit direction $(\theta=107,\phi=216)$ for the full range $\ell=2-591$ for this disk size. The two bins above $\ell=600$ however show a very different dipole direction. We see that the asymmetry can be seen as an alignment of the power distribution dipoles between multipoles from $\ell=2$ up to $\ell=600$. We will now study this alignment in more detail.

\begin{deluxetable}{ccc|ccccccc}
\tablecaption{  The table shows significances (in $\%$) and parameters $(\ell_\mathrm{max},\theta,\phi,A_0)$ of the best fit asymmetric 4-parameter model. Approximate Fisher matrix error for $\theta$, $\phi$ and $A_0$ are also given. Please refer to the text for details about the asymmetric models and their parameters. The significances specify the percentage of simulated maps with a larger drop in $\chi^2$ for the asymmetric model (considering only models with $\Delta\ell>200$) than found in the WMAP data.  The results are based on 1400 simulations.\label{tab:chires2}}
\tablewidth{0pt}
\tablehead{Channel&mask&$\ell$-range & {$\ell_\mathrm{max}$} & {$\theta$}(deg) & {$\phi$}(deg) & {p($\%$}) &  {$\Delta\theta$} & {$\Delta\phi$} & $A_0$ {$\times10^{-4}$ }}
\startdata
Q & Kq85 & $[2,300]$  & 221 & 104 & 226 & 0.1  & 10 & 10 & 1.6 $\pm$ 0.4 \\
V & Kq85 & $[2,300]$  & 221(281) & 107 & 226 & 0.3(0.3)$^*$ & 10 & 10 & 1.5 $\pm$ 0.4 \\
W & Kq85 & $[2,300]$  & 221 & 103 & 229 & 0.1  & 9 & 9 & 1.7 $\pm$ 0.4 \\
V & Kq75 & $[2,300]$  & 281 & 112 & 216 & 3.1 & 14 & 13 &  1.2 $\pm$ 0.4  \\
V & Kq75 ext. & $[2,300]$  & 300 & 114 & 202 & 1.5 & 12 & 13 & 1.3$\pm$0.4 \\
V & $|b|>30$ & $[2,300]$ & 300  & 131 & 170 & 17 & 18 & 19 & 1.0 $\pm$ 0.4 \\
\enddata\\
{\small $^*$ This results was obtained with 4200 simulations}
\end{deluxetable}

\begin{deluxetable}{ccc|cccccccc}
\tablecaption{ The table shows significances (in $\%$) and parameters $(\ell_\mathrm{max},\theta,\phi,A_0)$ of the best fit asymmetric 4-parameter model.  Approximate Fisher matrix error for $\theta$, $\phi$ and $A_0$ are also given. Please refer to the text for details about the asymmetric models and their parameters. The significances specify the percentage of simulated maps with a larger drop in $\chi^2$ for the asymmetric model than found in the WMAP data.  The results are based on 1400 simulations with the co-added V+W channels. \label{tab:chires3}}
\tablewidth{0pt}
\tablehead{Diameter&mask&$\ell$-range & {$\ell_\mathrm{max}$} & {$\theta$}(deg) & {$\phi$}(deg) & {p($\%$}) &  {$\Delta\theta$} & {$\Delta\phi$} & $A_0$ {$\times10^{-4}$}}
\startdata
{\bf $\Delta\ell>400$}  & & & & & & & & & & \\
\hline
$180^\circ$ & Kq85 & $[2,500]$  & 481 & 102 & 235 & 2.3 & 12 & 12 & 0.7 $\pm$ 0.2 \\
$180^\circ$ & Kq75 & $[2,500]$  & 481 & 104 & 224 & 14 & 16 & 16 & 0.5 $\pm$ 0.2  \\
$90^\circ$ & Kq85 & $[2,800]$  & 601 & 105 & 225 & 2.6 & 11 & 11 & 1.3$\pm$0.3 \\
$45^\circ$ & Kq85 & $[2,800]$  & 591 & 102 & 223 & 0.6 & 9 & 10 & 1.7$\pm$0.4 \\
$22.5^\circ$ & Kq85 & $[2,800]$  & 591 & 107 & 216 & 0.4 & 11 & 10 & 1.4$\pm$0.4\\
\hline
{\bf $\Delta\ell>200$}  & & & & & & & & & & \\
\hline
$90^\circ$ & Kq85 & $[2,300]$  & 221 & 102 & 229 & 0.3 & 11 & 11 & 2.6$\pm$0.7\\
$45^\circ$ & Kq85 & $[2,300]$  & 223 & 98 & 227 & 1.1 & 11 & 12 & 3.0$\pm$0.9\\
$22.5^\circ$ & Kq85 & $[2,300]$  & 223 & 97 & 220 & 2.1 & 14 & 14 & 0.21$\pm$0.07\\
\enddata\\

\end{deluxetable}

\begin{deluxetable}{ccc|ccccccccc}
\tablecaption{ The table shows significances (in $\%$) and parameters $(\ell_\mathrm{min},\ell_\mathrm{max},\theta,\phi,A_0)$ of the best fit asymmetric 5-parameter model.  Approximate Fisher matrix error for $\theta$, $\phi$ and $A_0$ are also given. Please refer to the text for details about the asymmetric models and their parameters. The significances specify the percentage of simulated maps with a larger drop in $\chi^2$ for the asymmetric model than found in the WMAP data.  The results are based on 1400 simulations for the co-added V+W channels. \label{tab:chires4}}
\tablewidth{0pt}
\tablehead{Diameter&mask&$\ell$-range & {$\ell_\mathrm{min}$} & {$\ell_\mathrm{max}$} & {$\theta$}(deg) & {$\phi$}(deg) & {p($\%$}) &  {$\Delta\theta$} & {$\Delta\phi$} & $A_0$ {$\times10^{-4}$}}
\startdata
{\bf $\Delta\ell>40$}  & & & & & & & & & & \\
\hline
$22.5^\circ$ & Kq85 & [2,95] & 2 & 63 & 110 & 226 & 8.4 & 15 & 16 & 0.38$\pm$0.14 \\
$22.5^\circ$ & Kq85 & [96,191] & 96 & 191 & 100 & 200 & 19 & 16 & 16 & 0.27$\pm$0.10 \\
$22.5^\circ$ & kq85 & [192,303] & 208 & 281 & 100 & 238 & 99 & 35 & 34 & 0.13$\pm$0.11\\
$22.5^\circ$ & kq85 & [304,399] & 352 & 399 & 83 & 182 & 77 & 24 & 24 & 0.23$\pm$0.13 \\
$22.5^\circ$ & kq85 & [400,495] & 432 & 479 & 113 & 224 & 36 & 19 & 19 & 0.32$\pm$0.14\\
$22.5^\circ$ & kq85 & [496,591] & 496 & 591 & 112 & 210 & 48 & 20 & 20 & 0.21$\pm$0.10\\
$22.5^\circ$ & kq85 & [592,703] & 608 & 687 & 45 & 111 & 43 & 20 & 24 & 0.26$\pm$0.11 \\
$22.5^\circ$ & kq85 & [704,799] & 736 & 799 & 35 & 47 & 63 & 26 & 36 & 0.25$\pm$0.13 \\
\enddata\\

\end{deluxetable}


\subsection{Results with the alignment test}

\begin{figure}
{\vspace*{-8cm}\hspace*{-5cm}\includegraphics[width=1.6\linewidth]{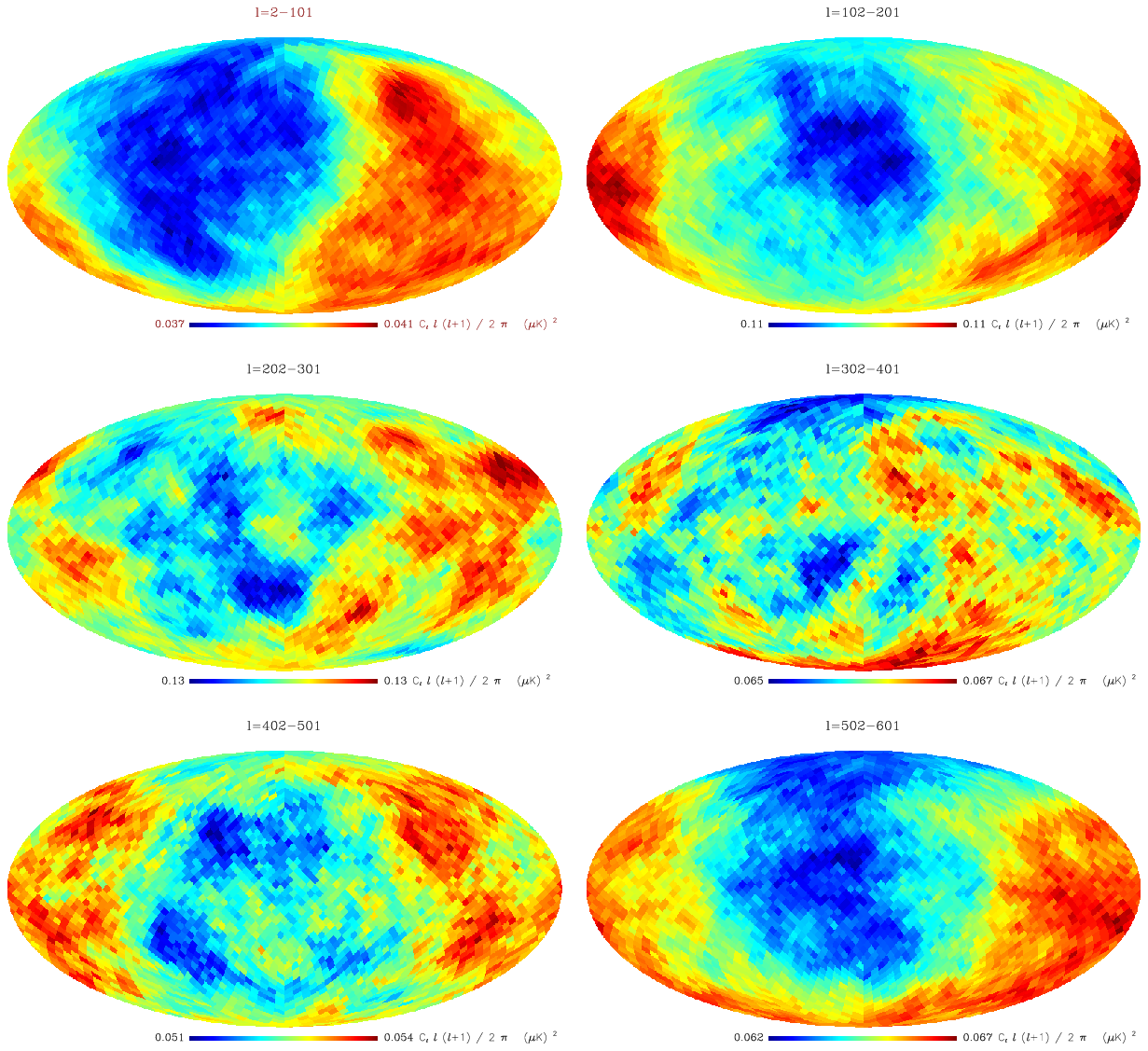}}
\vspace*{-12cm}\caption{The distribution of power in blocks of 100 multipoles estimated on hemispheres for the combined V+W band using the KQ85 sky cut. Note the similarity with the single multipole bins in figure \ref{fig:outliers}\label{fig:directions}}
\end{figure}

\begin{figure}
\begin{center}
\includegraphics[angle=90,width=0.7\linewidth]{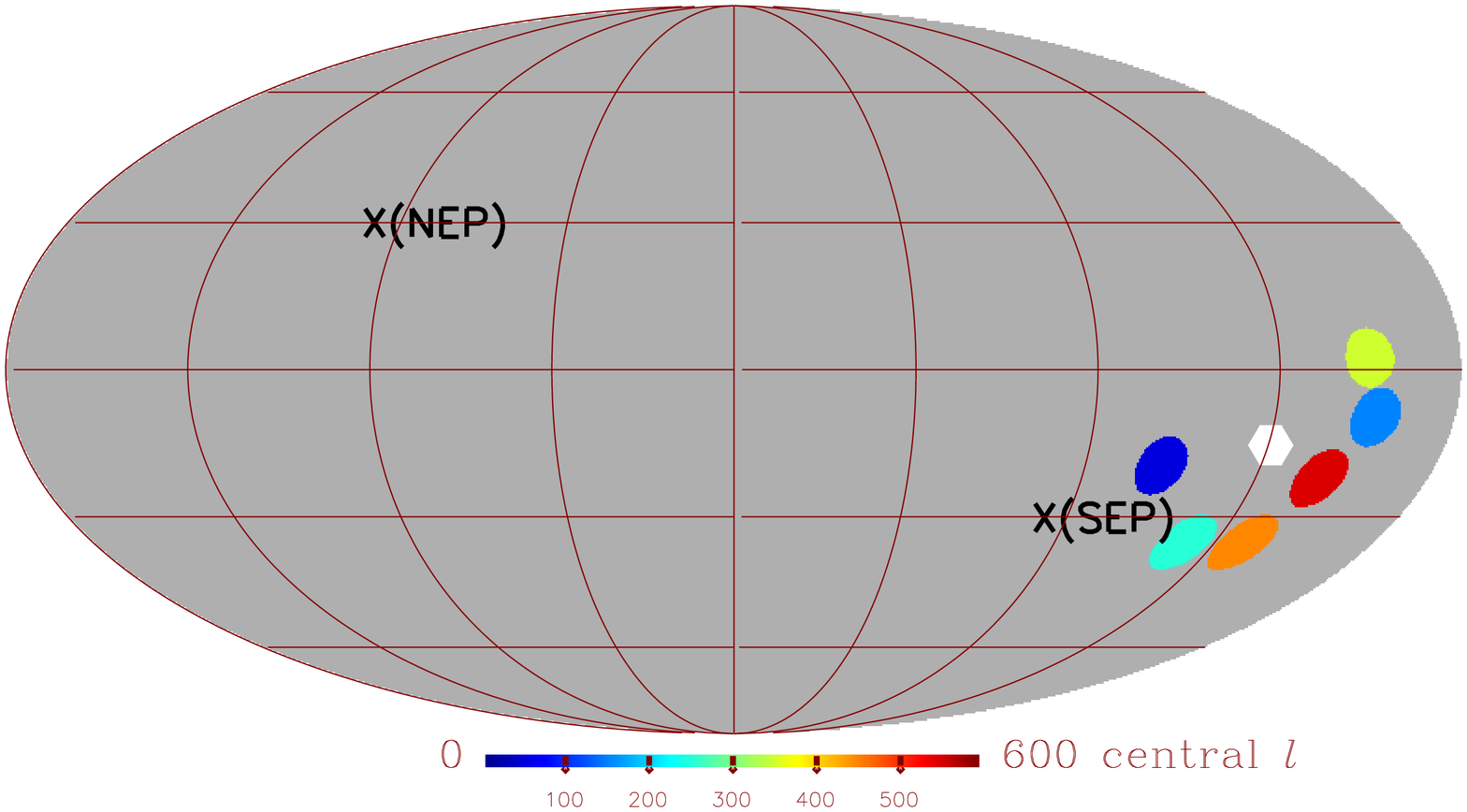}
\caption{The directions of the dipoles of the power distribution in blocks of 100 multipoles estimated on disks with diameter of 22.5 degree for the combined V+W band using the KQ85 sky cut.  The colour of the disks indicate the center of the given multipole range.  The white hexagon indicates the best fit dipole direction for the full range $\ell=2-600$. The ecliptic poles are indicated by crosses.\label{fig:directions3}}
\end{center}
\end{figure}

Before looking at significances, we will illustrate the direction of the dipoles of individual 100 multipole blocks with some figures. In figure \ref{fig:directions} we show the distribution of power in the WMAP V+W band data using hemispheres (KQ85 cut was used in the power spectrum estimation). Each map shows the distribution of power $M_i(b1,b2)$ for a given 100-multipole range. We see already by eye that there is a clear dipolar distribution and that the direction of the dipole is very similar in each case. In figure \ref{fig:directions3} we show the position of the dipole for subranges of 100 multipoles. The color of the disk indicates the multipole range (see table \ref{tab:chires4} for the exact ranges used). The results in this plot is taken from power spectrum estimates on disks with diameter 22.5 degree using the KQ85 galactic cut. We see that all the individual multipole ranges have dipoles pointing in a direction close to the best fit dipole for the full range $\ell=2-600$ indicated by the white hexagon.

\begin{figure}
\begin{center}
\vspace*{-10cm}\hspace*{-5cm}\includegraphics[width=1.6\linewidth]{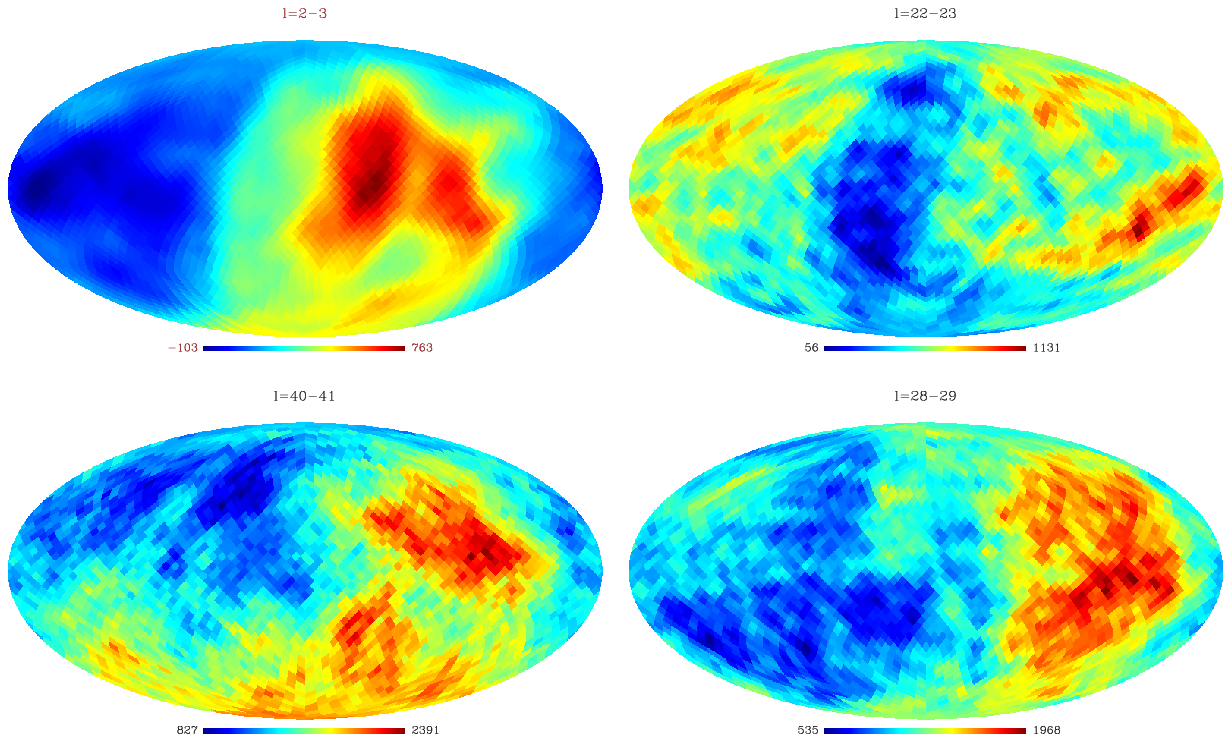}
\vspace*{-15cm}\caption{The distribution of power $\ell(\ell+1)C_\ell$ for the multipole ranges $\ell=2-3$ (upper left), $\ell=22-23$ (upper right), $\ell=40-41$ (lower left) and $\ell=28-29$ (lower right)   \label{fig:outliers}}
\end{center}
\end{figure}

In \cite{asymm2}, we point out that the outliers in the full sky spectrum at $\ell=22$ and $\ell=40$ seem to be associated with the asymmetry:  The high outlier at $\ell=40$ was associated with the high power in the hemisphere of maximum asymmetry and the low outlier at $\ell=22$ was associated with the low power in the opposite hemisphere and. In figure \ref{fig:outliers} we show the distribution of power in these two bins as well as for the first bin $\ell=2-3$ and the bin $\ell=28-29$ which is a particularly asymmetric bin. The conclusions from \cite{asymm2} still hold. One can see that the same dipolar distribution of power which is also seen in the 6 subranges between $\ell=2-600$ in figure \ref{fig:directions}.

In table \ref{tab:blocks20} we show the results for the alignment test using blocks of 20 multipoles. We limit the maximum multipole to 300 because noise is getting important after $\ell=300$ increasing the variance of the directions of blocks with just 20 multipoles. Later for blocks of 100 multipoles we will also consider higher multipoles as the noise is reduced in each block when averaging over 100 instead of 20 multipoles. In the table we have considered the alignment test for the V band using different galactic masks. The numbers indicate the percentage of simulations with a lower mean angle $\bar\theta$ between dipole directions. For $\ell=2-300$ there is a significant ($\sim1\%$ level) alignment between dipole directions for the 20 multipole blocks. Note in particular that the for the extended KQ75 cut, none of the 1400 simulations are as strongly aligned as the WMAP data. Thus, the asymmetry is strong even with a large galactic cut. We also see that there is a significant ($2-4\%$ level) alignment when considering only the first five 20-multipole blocks on large scales $\ell=2-100$. Considering only the blocks on small scales $\ell=200-300$ excluding the first 100 multipoles however, there is no significant alignment present.

In table \ref{tab:blocks100_2} we show the results from the alignment test of 100 multipole blocks using more localized spectra. The V+W map with the KQ85 galactic cut was used in this analysis. We clearly see the result of the strong alignment which was already obvious in table \ref{tab:chires4}. Using more localized spectra the alignment appears even stronger than for hemisphere spectra. Note that for the range $\ell=2-600$ none of the simulations show a similarly strong alignment for any of the disk sizes. In particular for the $22.5^\circ$ (diameter) disk results, none of 9800 simulations have a similarly strong alignment which is close to a $4\sigma$ detection of asymmetry. Note also that the alignment is highly significant also for separate multipole ranges at small and large scales, for instance $\ell=2-300$ and $\ell=300-600$.

\begin{deluxetable}{c|cccc}
\tablecaption{For each entry in the table we have calculated the mean angle $\bar\theta$ between the dipole directions for all blocks of 20 multipoles within the given multipole range (refer to the text for details of how $\bar\theta$ is calculated). The numbers given in the table is the percentage of simulated maps with a lower mean angle $\bar\theta$. The results are based on 1400 simulations. Zero entries means that none of the simulated maps had a similarly low mean angle.\label{tab:blocks20}}
\tablewidth{0pt}
\tablehead{
mask:&KQ85(V)&KQ75(V)&KQ75ext(V)&$|b|>30$(V)}
\startdata
\hline
$\ell=2-300$ & 0.9  & 0.9 & 0 & 5.6 \\
$\ell=2-200$ & 1.8  & 5.5 & 0.8 & 13 \\
$\ell=2-100$ & 2.0  & 4.3 & 3.4 & 16 \\
$\ell=100-300$ & 9.0  & 7.4 & 1.8 & 35.5 \\
$\ell=200-300$ & 67.5  & 24.6 & 20 & 49.9 \\
\enddata
\end{deluxetable}

\begin{deluxetable}{c|ccccc}
\tablecaption{For each entry in the table we have calculated the mean angle $\bar\theta$ between the dipole directions for all blocks of 100 multipoles within the given multipole range (refer to the text for details of how $\bar\theta$ is calculated). The numbers given in the table is the percentage of simulated maps with a lower mean angle $\bar\theta$. The results are based on 1400 simulations except for the results for $22.5^\circ$ disks which is based on 9800 simulations. Zero entries means that none of the simulated maps had a similarly low mean angle.  The combined V+W map was used in obtaining all results in this table. The KQ85 mask was used when other mask is not specified. \label{tab:blocks100_2}}
\tablewidth{0pt}
\tablehead{
disk size:&$180^\circ$&$90^\circ$&$45^\circ$&$22.5^\circ$&$90^\circ$(KQ85N)}
\startdata
\hline
$\ell=2-800$     &   & 6.1   & 8.3 & 9.2 & 53 \\
$\ell=2-700$     &   & 0.5   & 0.5 & 0.4 & 15 \\
$\ell=2-600$     &   & 0    & 0   & 0 & 1.1 \\
$\ell=2-500$     & 0.1   & 0  & 0   & 0.04 & 3.9 \\
$\ell=2-400$     & 0.1   & 0  & 0.1 & 0.3 & 4.6\\
$\ell=2-300$     & 0.5   & 0.6  & 0.7 & 1.8 & 11 \\
$\ell=2-200$     & 14   & 13 & 9.2 & 18 & 41\\
$\ell=200-600$ & & 0  & 0   & 0  & 2.7 \\
$\ell=300-600$ & & 0.1  & 0.1 & 0.1 & 7.3 \\
$\ell=400-600$ & & 3.6   & 0.4 & 0.5 & 6.6\\
\enddata
\end{deluxetable}

\begin{figure}
\begin{center}
\vspace*{-10cm}\hspace*{-5cm}\includegraphics[width=1.6\linewidth]{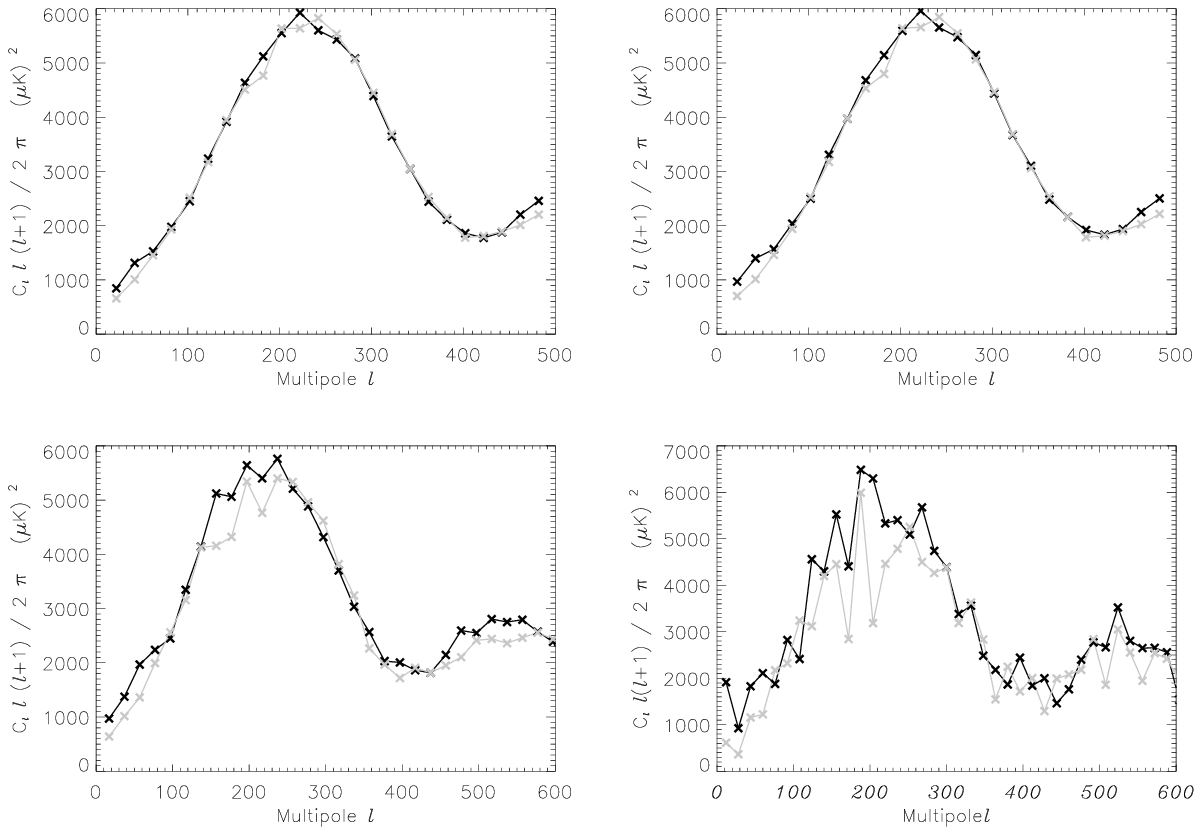}
\vspace*{-15cm}\caption{ Power spectra obtained in opposite directions. The upper two plots are spectra estimated on hemispheres, one centered at $(102^\circ,235^\circ)$ (black line) and one in the opposite direction (grey line). The left plot shows the spectra after foreground subtraction (by the WMAP team), the right plot shows the same spectra before foreground subtraction. We see that if the asymmetry is larger than the foreground correction which is applied to the maps. The lower two plots show the more localized spectra taken in opposite directions, left plot for spectra estimated on 90 degree (diameter) disks, right plot for 45 degree disks. The asymmetry is more pronounced for more localized spectra. \label{fig:cl}}

\end{center}
\end{figure}

\begin{figure}
\begin{center}
\vspace*{-8cm}\hspace*{-5cm}\includegraphics[width=1.6\linewidth]{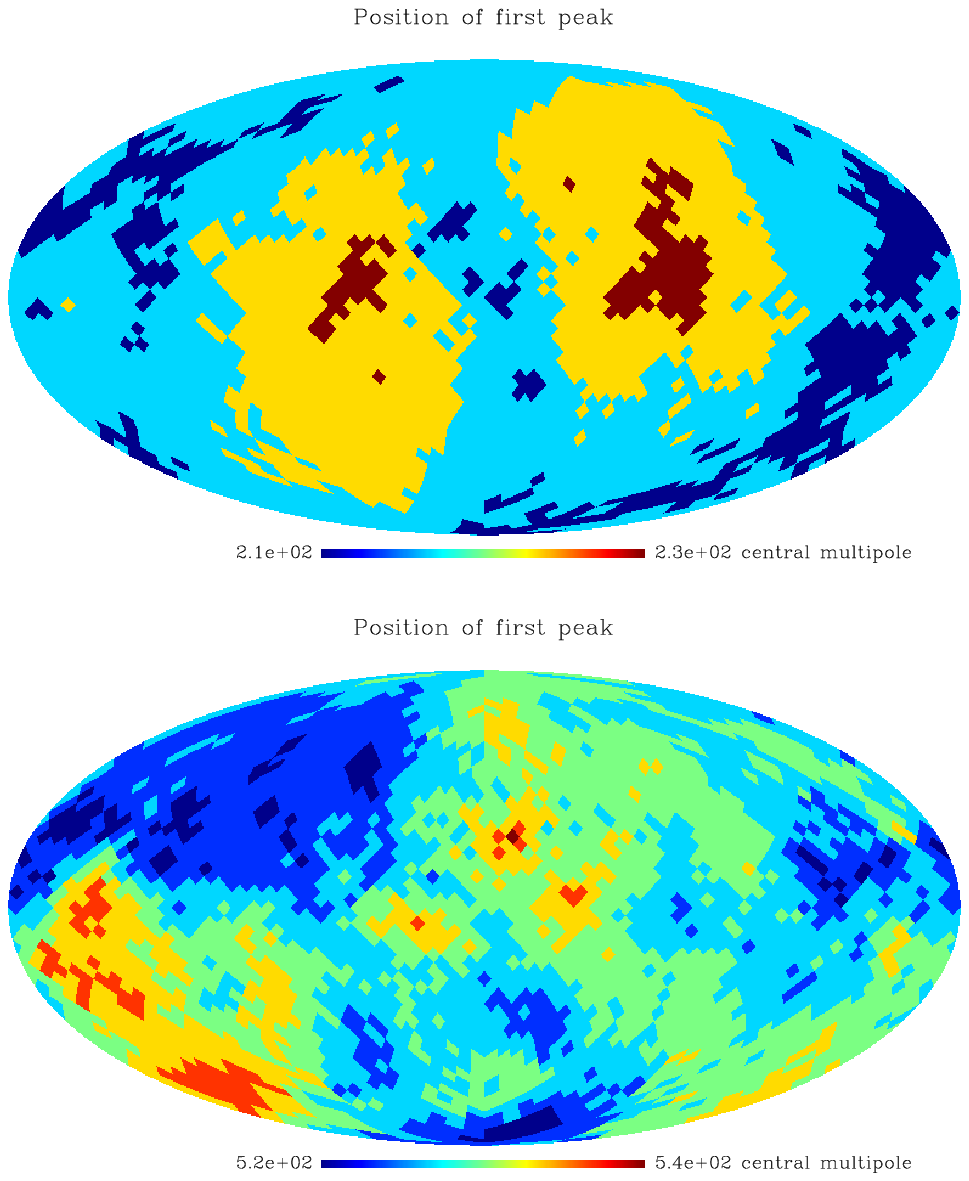}
\vspace*{-12cm}\caption{ For each position on the sphere, the colour indicates the multipole positions of the first (upper plot) and the second (lower plot) Doppler peak. The spectra were estimated on disks with diameter of $90^\circ$ centered on the given position on the sphere. The combined V+W map with the KQ85 sky cut was used for the power spectrum estimation.\label{fig:peak}}
\end{center}
\end{figure}

In figure \ref{fig:cl} we show the spectra in the best fit dipole direction for $\ell=2-600$ for various disk sizes from hemispheres to $45^\circ$ (diameter) disks. Also in these plots we see that the difference between the spectra in the opposite directions becomes larger with more localized spectra. In particular the first part of the spectrum $\ell=2-100$ as well as the amplitudes of the first two peaks are clearly different. We have investigated whether also the positions and not only the amplitudes of the first two peaks may have a similarly asymmetric distribution on the sky. For the localized spectra obtained on $90^\circ$ (diameter) disks using the V+W map with the KQ85 cut, we made a fit to the first two peaks. In figure \ref{fig:peak} we show the distribution of the multipole position of the first and second peak. No dipole structure similar to what was found for the power spectrum amplitude is seen in these figures.

\subsection{Testing foregrounds and systematics}

In this section we will perform several tests in order to investigate whether foreground residuals or instrumental systematic effects may cause the observed asymmetry. In particular, we will look at the cross spectra instead of the auto-spectra, we will look at the WMAP data year by year and finally we will study in detail whether there are still clear signs of the asymmetry outside the $|b|>30$ cut.

The cross spectra based on $a_{\ell m}$ obtained from different channels and years of observations are less prone to systematical errors and in particular to uncertainties in the noise model \citep{hinshaw}. We have obtain the hemisphere spectra based on the spectrum obtained as a mean of all 780 possible combinations of the channels Q, V and W as well as the five years of observation. Making simulations with all 780 cross-spectra turned out to require too much CPU time and we were therefore not able to perform a full statistical test using the cross-spectra. Using the WMAP data alone we found that the direction of the dipole using the maps $M_i^b$ based on cross-spectra is consistent with the dipole based on the auto-spectrum.

Similarly we have studied the direction of the dipole for each single year of observation. We found that these directions are consistent with the direction obtained with the co-added maps with all years included. There is thus no sign of systematic errors in specific years causing the asymmetry. We have also considered the difference between the power distribution maps $M_i(b1,b2)$ obtained with different channels. Foreground residuals causing the asymmetry would show up in these differences between channels. The dipole directions of the two difference maps Q-V and V-W are not similar to the direction of asymmetry. There is thus no sign of a frequency dependent foreground with the dipolar power distribution which we have detected in the individual bands.

As discussed above, the asymmetry is no longer significant for the $|b|>30$ cut, but the fact that the direction of asymmetry is still (within the error bars) consistent with the direction of asymmetry found for smaller galactic cuts is a strong argument against foreground residuals causing the asymmetry.

Still, in order to make sure that the drop in significance for the 60 degree cut is not due to the fact that the large mask is excluding some galactic residual causing the asymmetry, we made some further tests. First we made a mask which was equal to the regular KQ85 cut for the southern galactic hemisphere, but had an extended 30 degree galactic cut in the northern galactic hemisphere (this is the KQ85N mask). We also made a similar mask extending only in the southern hemisphere (this is the KQ85S mask). 

Using the southern KQ85S mask (for $\ell=2-300$), the significance of the asymmetry is still high ($p=1.2\%$) for the range $\ell=2-261$ (the direction of asymmetry is consistent with the results above). For the northern KQ85N mask the significance has dropped to $12\%$, but the direction of asymmetry is still consistent. Using the more localized spectra estimated on $90^\circ$ disks, we see from table \ref{tab:blocks100_2} that the alignment between 100-multipole blocks is  significant at the $1.1\%$ level for $\ell=2-600$ using the KQ85N cut. For $\ell=2-400$ the alignment is still significant at the $2\sigma$ level. This result combined with the fact that the direction of asymmetry has changed little with the large $|b|>30$ cut as well as the consistency of results using different frequency bands shows that an explanation of the asymmetry in terms of  foreground residuals is difficult to make consistent with the results presented in this paper.

\subsection{Non-flat amplitude}

\begin{figure}
\begin{center}
\includegraphics[width=0.45\linewidth]{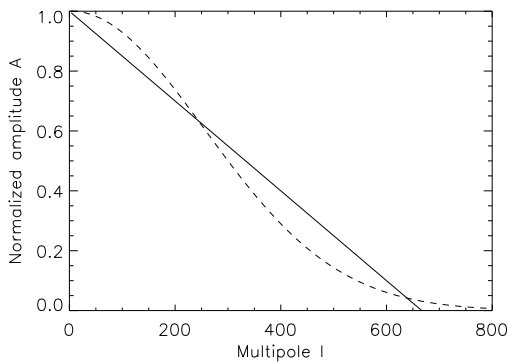}
\caption{The multipole dependence of the amplitude of  asymmetry A. We show the results of the fit to a linear and Gaussian model of A. The results were obtained from spectra estimated on 90 degree (diameter) disks using the V+W band with the KQ85 sky cut. \label{fig:multipoledep}}
\end{center}
\end{figure}

We have so far discussed a model where the common dipole has the same amplitude $A_0$ over the full multipole range in the best fit model. We have also tested a model with  $A$ decreasing linearly with multipole as well as a Gaussian shape of the amplitude. Both these models are described in detail in section \ref{sect:met2}. In table \ref{tab:chiresvar} we present the results. The first part of the table shows the result with the linear fit where $\alpha$ is the parameter describing how fast $A$ decreases with multipole. We see that a model with decreasing amplitude is preferred by the data. In figure \ref{fig:multipoledep} we show the multipole dependence of the asymmetry. The asymmetry decreases and vanishes close to $\ell=600$ consistent with the results in the previous chapters.

In the lower part of the table, the results with the Gaussian model are shown. We shows the results for the 4-parameter model where the peak of the Gaussian is forced to $\ell_\mathrm{peak}=2$ with the width $\sigma$ allowed to vary as well as for the 5-parameter model with $\ell_\mathrm{peak}$ as an additional free parameter. With the exception of the 45 degree disk results, the 5 parameter model finds the same best fit model with $\ell_\mathrm{peak}=2$ as the 4-parameter model. In figure  \ref{fig:multipoledep} we have plotted the best fit Gaussian model for 90 degree disks on top of the linear model. We see that the two models show a consistent decrease in the amplitude of the asymmetry. We conclude that the asymmetry is larger for smaller multipoles and decreasing continuously towards $\ell=600$ where it disappears.

\begin{deluxetable}{cc|ccccc}
\tablecaption{   Significances and directions of the common dipole component for asymmetric models with linearly decreasing or Gaussian amplitude profile. We show the significances (in $\%$) and parameters $(\theta,\phi)$ as well as $\alpha$ for the linear model. For the Gaussian model we show the best fit parameters  $\sigma$ (included only in the 5-parameter fit) and  $\ell_\mathrm{peak}$. Please refer to the text for details about the asymmetric models and their parameters. The significances specify the percentage of simulated maps with a larger drop in $\chi^2$ for the asymmetric model than found in the WMAP data. The results are based on 1400 simulations. \label{tab:chiresvar}}
\tablewidth{0pt}
\tablehead{Channel& Mask & {$\alpha$/$\ell_\mathrm{peak}$} & {$\theta$} (deg)& {$\phi$} (deg) & $\sigma$ & {p(\%)}}
\startdata
{\bf Linear}  & & & & &  \\
\hline
V+W & KQ85  & 0.0018 & 99 & 229 & &0.6 \\
V+W & KQ75  & 0.0018 & 100 & 218 & & 17 \\
V+W & KQ85 ($90^\circ$ disks) & 0.0015 & 100 & 230 & & 1.3 \\
V+W & KQ85 ($45^\circ$ disks) & 0.0014 & 99 & 228 & & 1.4 \\
\hline
{\bf Gauss}  & & & & & \\
\hline
V+W & KQ85  & NA & 230 & 100 & 228 & 0.4 \\
V+W & KQ85  & 2 & 230 & 100 & 228 & 1.7 \\
V+W & KQ85 ($90^\circ$ disks)  & NA & 252 & 100 & 230 & 0.8 \\
V+W & KQ85 ($90^\circ$ disks)  & 2 & 252 & 100 & 230 & 14 \\
V+W & KQ85 ($45^\circ$ disks)  & NA & 312 & 100 & 228 & 1.4 \\
V+W & KQ85 ($45^\circ$ disks) & 408 & 312 & 100 & 228 & 9.1 \\
\enddata
\end{deluxetable}

\section{Conclusions}
\label{sect:conclusions}

We have reassessed the asymmetry in the distribution of CMB fluctuation  power on the sky reported in \cite{asymm1} and \cite{asymm2}. In order to test whether an anisotropic model of the CMB fluctuations is actually preferred over an isotropic model taking into account the additional parameters required, we implement a new model selection procedure. We model the asymmetric distribution of power on the sky as a dipole in the power distribution. Note that this is not to say that the CMB fluctuation field has a dipole, rather that the power for a certain scale (multipole) has a dipole distribution on the sky. We use a model where there is a common dipole component in the power distribution for a set of multipoles in a range $[\ell_\mathrm{min},\ell_\mathrm{max}]$ where $\ell_\mathrm{min}$ and $\ell_\mathrm{max}$ are free parameters as well as the direction ($\theta,\phi$) and the amplitude $A_0$ of the dipole. We use a $\chi^2$ approach to find the best fit model parameters among these 5 parameters.

We first investigated a model where we assume the asymmetry to start at $\ell_\mathrm{min}=2$, reducing the number of free parameters in the model to 4. Using power spectra estimated on hemispheres in the combined V+W band with the KQ85 galactic cut,  we find a strong asymmetry in the multipole range $\ell=2-221$ with an axis pointing in the direction $(\theta=107^\circ\pm10^\circ,\phi=226^\circ\pm10^\circ)$ (which is the direction where the power is largest). Only $0.3\%$ of the simulated isotropic maps show a similarly strong asymmetry. Performing the same test on the Q, V and W bands individually as well as with different galactic cuts, a similar asymmetry is found. The significance is reduced for larger cuts but for an extended KQ75 cut (excluding $37\%$ of the sky), only $1.5\%$ of the simulated isotropic map show a simiarly strong common dipole component in the range $\ell=2-300$. Using more localized power spectra estimated on smaller disks, we perform the same tests on the V+W band using the KQ85 cut. Smaller disks allow faster spherical harmonic transforms and allows the analysis to include multipoles up to $\ell=800$. We find that the range $\ell=2-600$ is asymmetric with dipole direction  $(\theta=107^\circ\pm11^\circ,\phi=216^\circ\pm10^\circ)$ using the smallest $22.5^\circ$ diameter disks. Only $0.4\%$ of the simulated maps show similar asymmetry. In figure \ref{fig:cl} we show the spectra in the two opposite parts of the sky. The spectra are clearly different for the largest scales as well as around the two first peaks.

Including $\ell_\mathrm{min}$ as a free parameter, the same anisotropic model is favored and the best fit value for the first multipole with a common dipole is $\ell_\mathrm{min}=2$. We therefore concluded that the 4 parameter model was sufficient to describe the asymmetry.  Testing models with a multipole dependent dipole amplitude, we found that a model where the asymmetry is maximum for small $\ell$ and decreasing with increasing $\ell$ vanishing at about $\ell\sim650$ gives a good fit to the data ($0.4\%$).

To check whether the asymmetry is present in the full range $\ell=2-600$ or only for some multipoles, we performed a second simpler test of asymmetry. We found the dipole direction for the power distribution in multipole ranges of 100 multipoles, $\ell=2-101$, $\ell=102-201$, etc. to $\ell=502-601$. We thus obtained 6 dipole directions from 6 independent multipole ranges. The power distribution in these ranges are shown in figure \ref{fig:directions}. We found that these 6 dipoles were much more aligned in the WMAP data than in isotropic simulations. In fact, using power spectra estimated on $22.5^\circ$ diameter disks from the combined V+W bands with KQ85 galactic cut we found that none of our 9800 simulations show a similarly strong alignment between these 6 dipoles. The dipole directions for these 6 ranges are shown in figure \ref{fig:directions3}. We find that the spatial distribution of CMB fluctuations is strongly correlated between small and large angular scales.

The fact that all the frequency channels,  all years of observations and also tests using cross power spectra show a similarly asymmetric distribution strongly disfavors an explanation in terms of systematic effects and residual galactic foregrounds. Further, in the 5 year WMAP data a different approach to foreground subtraction than for the first year data was applied, still the asymmetry remained significant at the same level for the large scale asymmetry $\ell=2-40$. In \citep{wifit} we also showed that a blind approach to foreground subtraction did not change this result for large scales.

Our results indicate that the reported common asymmetric axis extending over a large range in scales is highly unlikely to be a statistical fluke. Foregrounds and systematic effects do not seem to be probable explanations. The CMB does seem to have an uneven power distribution on the sky over a large range of angular scales. An important task for further research is to find a physical explanation for this asymmetry which can predict possible effects on CMB polarization to be tested in future experiments.

\begin{acknowledgments}
  We acknowledge the use of the HEALPix \citep{healpix} package. FKH is thankful for an OYI grant from the Research Council of Norway.  We acknowledge the use of the NOTUR super computing facilities. We acknowledge the use of the Legacy Archive for Microwave Background Data Analysis (LAMBDA). Support for LAMBDA is provided by the NASA Office of Space Science.
\end{acknowledgments}

\end{document}